\documentclass[aps,prl,preprint,amsmath,amssymb,superscriptaddress]{revtex4-1}
\usepackage{graphicx}
\usepackage{color}
\usepackage{hyperref}

\begin{document}

\title{Magnetic proximity effect in $\mathbf{YBa_2Cu_3O_7/La_{2/3}Ca_{1/3}MnO_3}$ and  $\mathbf{YBa_2Cu_3O_7/LaMnO_{3+\delta}}$ superlattices}

\author{D.\,K. Satapathy}
\author{M.\,A. \surname{Uribe-Laverde}}
\author{I. Marozau}
\author{V.\,K. Malik}
\author{S. Das}
\author{Th.~Wagner}
\affiliation{University of Fribourg, Department of Physics and Fribourg Centre for Nanomaterials, Chemin du Mus\'ee 3, CH-1700 Fribourg, Switzerland}
\author{C. Marcelot}
\author{J.~Stahn}
\affiliation{Laboratory for Neutron Scattering Paul Scherrer Institut, CH-5232 Villigen, Switzerland}
\author{S. Br\"uck}
\affiliation{University of W\"urzburg, Physikalisches Institut, Am Hubland, D-97074 W\"urzburg, Germany}
\author{A. R\"uhm}
\author{S. Macke}
\author{T. Tietze}
\author{E. Goering}
\affiliation{Max-Planck-Institut f\"ur Intelligente Systeme, Heisenbergstrasse 3, D-70569 Stuttgart, Germany}

\author{A. Fra\~n\'o}
\affiliation{Max-Planck-Institut f\"ur Festk\"orperforschung, Heisenbergstrasse 1, D-70569 Stuttgart, Germany}
\affiliation{Helmholtz-Zentrum Berlin f\"ur Materialien und Energie, Wilhelm-Conrad-R\"ontgen-Campus BESSY II, Albert-Einstein-Str. 15, D-12489 Berlin, Germany}

\author{J.\,-H. Kim}
\author{M. Wu}
\author{E.~Benckiser}
\author{B. Keimer}

\affiliation{Max-Planck-Institut f\"ur Festk\"orperforschung, Heisenbergstrasse 1, D-70569 Stuttgart, Germany}

\author{A. Devishvili}
\author{B.\,P. Toperverg}
\affiliation{Institute of Solid State Physics, Ruhr-Universit\"at Bochum,  D-44801 Bochum, Germany}

\author{M. Merz}
\author{P. Nagel}
\author{S. Schuppler}
\affiliation{Karlsruhe Institute of Technology, Institut f\"ur Festk\"orperphysik, D-76021 Karlsruhe, Germany}

\author{C. Bernhard}
\affiliation{University of Fribourg, Department of Physics and Fribourg Centre for Nanomaterials, Chemin du Mus\'ee 3, CH-1700 Fribourg, Switzerland}


\date{\today}

\begin{abstract}
Using neutron reflectometry and resonant x-ray techniques we studied the magnetic proximity effect (MPE) in superlattices composed of superconducting YBa$_2$Cu$_3$O$_7$~(YBCO) and ferromagnetic-metallic (FM-M)  La$_{0.67}$Ca$_{0.33}$MnO$_{3}$~(LCMO) or ferromagnetic-insulating (FM-I) LaMnO$_{3+\delta}$~(LMO). We find that the MPE strongly depends on the electronic state of the manganite layers, being pronounced for the FM-M LCMO and almost absent for FM-I LMO. We also detail the change of the magnetic depth profile due to the MPE and provide evidence for its intrinsic nature.
\end{abstract}

%\keywords{}

\maketitle

The coupling between the antagonistic superconducting (SC) and ferromagnetic (FM) orders in thin film multilayers is the subject of intensive research. The understanding of the SC/FM multilayers comprising conventional elemental or alloy materials is already fairly advanced~\cite{Buzdin_2004,Eschrig_2011}. In comparison, little is known about their oxide counterparts which combine the cuprate high $T_\text{c}$ superconductors (HTSC) and the manganites that are well known for their colossal magnetoresistance (CMR) effect~\cite{Helmolt_1993,Tokura_1999}. A few groups have been growing  YBa$_2$Cu$_3$O$_7$/La$_{2/3}$Ca$_{1/3}$MnO$_{3}$ (YBCO/LCMO) multilayers~\cite{Przyslupski_1997,Goldman_2001,Habermeier_2001,Sefrioui_2003,Holden_2004,Werner_2010} and observed a range of interesting effects, like a giant magnetoresistance~\cite{Pena_2005}, a large photo-induced $T_{\text{c}}$ increase~\cite{Pena_2006}, a magnetic proximity effect (MPE)~\cite{Stahn_2005,Hoffmann_2005,Chakhalian_2006,Hoppler_2009} where a FM Cu moment is induced in the YBCO layers, a reconstruction of the orbital occupation and symmetry of the interfacial CuO$_2$ layers~\cite{Chakhalian_2007} and even a SC-induced modification of the FM order~\cite{Hoppler_2009}. These phenomena are thought to be closely related to the strong electronic correlations and the intimate coupling between the magnetic, orbital and lattice degrees of freedom in these oxides. These yield a manifold of nearly degenerate ground states with a rich spectrum of properties that can be readily tuned with external parameters. The cuprate HTSC are indeed close to a magnetic instability where a spin-density-wave (SDW) is induced with impurities~\cite{Alloul_2009,Suchanek_2010} or magnetic fields~\cite{Lake_2002}. The manganites have also extremely versatile electromagnetic properties that are strongly modified by strain, pressure or magnetic field~\cite{Tokura_1999}. It is thus not unexpected that the proximity effect in these oxide SC/FM superlattices (SLs) is more complex than in their classical counterparts and involves an unusual magnetic component that remains to be understood. 

Here we establish that the electronic and/or orbital properties of the manganite layers are playing a decisive role in the MPE. We show that the MPE is almost absent in SLs with FM-I LaMnO$_{3+\delta}$ (LMO) layers and we detail how the MPE affects the FM depth profile for a SL with FM-M LCMO layers.

Superlattices of [YBCO(10\,nm)/LCMO(10\,nm)]$_{10}$ and  [YBCO(10nm)/LMO(10\,nm)]$_{10}$ were grown with pulsed laser deposition (PLD) on La$_{0.3}$Sr$_{0.7}$Al$_{0.65}$Ta$_{0.35}$O$_3$~(LSAT) (001) substrates (10$\times$10$\times$0.5\,mm$^3$) and characterized as described in Ref.~\cite{Malik}. The SC onset temperatures are $T_{\text{c}}^{\text{ons}}\approx 88$\,K and 77\,K and the FM ones are at $T^{\text{Curie}}\approx200$\,K and 140\,K, respectively. As outlined in Ref.~\cite{EPAPS} the LMO layers are non stoichiometric and thus hole doped to a FM-I state~\cite{Marton_2010}. 
 
Polarized neutron reflectometry (PNR) measurements were performed with the two-axis diffractometers AMOR at SINQ of PSI in Villigen, CH; NREX of FRM-II in Munich, D; and SUPERADAM of ILL in Grenoble, F.  Magnetic fields up to 4\,kOe oriented perpendicular to the scattering plane and parallel to the film surface were produced with Helmholtz coils. The samples were cooled with closed-cycle cryostats in applied magnetic field. Data fitting was performed with the program Superfit~\cite{Superfit}. The x-ray magnetic circular dichroism (XMCD) was measured at the beamlines UE56/2-PGM1 at BESSY II, Helmholtz Center Berlin, D; and WERA at ANKA in Karlsruhe, D. The  x-ray resonant magnetic reflectometry (XRMR) was measured on UE56/2-PGM1 using the MPI-IS ErNST endstation. Data simulation was performed with the program ReMagX~\cite{Remagx}. Details about the XRMR technique are given in Refs.~\cite{Brueck_2008,EPAPS}. 
\begin{figure}
\includegraphics[width=0.7\textwidth]{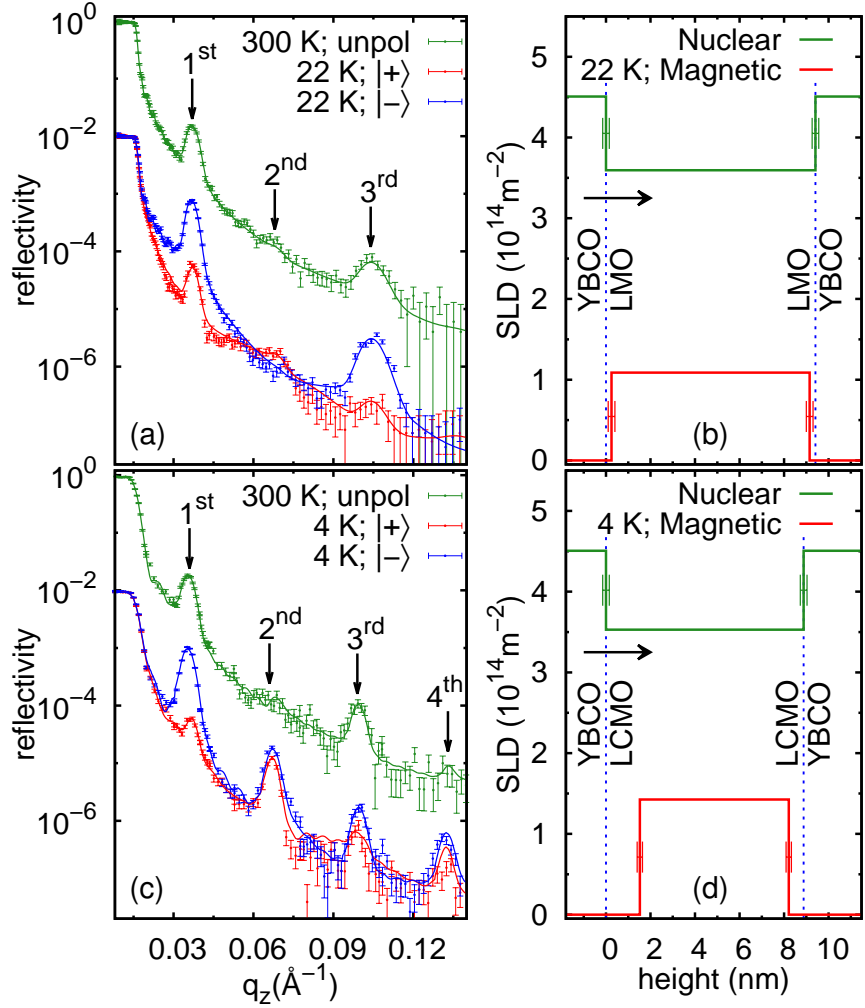}
\caption{PNR curves for spin-up ($|$+$\rangle$) and spin-down ($|\text{-}\rangle$) polarization of \textbf{a)} [YBCO(10nm)/LMO(10nm)]$_{10}$ and \textbf{c)} [YBCO(10nm)/LCMO(10nm)]$_{10}$. Also shown are unpolarized curves at 300K. Arrows mark the position of the SLBPs. The curves are vertically shifted for clarity. \textbf{b)} and \textbf{d)}, depth profiles as obtained from fitting the data in \textbf{a)} and \textbf{c)} (solid lines) of the nuclear (green line) and magnetic (red line)  scattering length densities (SLD) which are proportional to the nuclear and magnetic potentials, respectively. Arrows point along the SL growth direction. The error for the SLD in \textbf{b)} and \textbf{d)} is smaller than the line thickness, all the error bars in these two figures were obtained as described in Ref.~\cite{EPAPS}.\label{Fig1}}
\end{figure}

Figure \ref{Fig1} displays our specular PNR data on the YBCO/LCMO and YBCO/LMO SLs which establish that the MPE  strongly depends on the electronic properties of the FM layers. Figures ~\ref{Fig1}a and~\ref{Fig1}c compare the PNR curves at $T\ll T^{\text{Curie}}$. They confirm that the LMO and the LCMO layers are both strongly FM. This is evident from the large splitting of the spin-up and spin-down polarized curves around the 1$^\text{st}$ order superlattice Bragg peak (SLBP) which is a measure of the average magnetization density of the SL. The best fits (solid lines)  with a simple model of block-like potentials yield an average moment per Mn ion of 2.3\,$\mu_B$ for both LMO and LCMO, that agrees well with the saturation value of 2.4 and 2.3\,$\mu_B$ from dc magnetization (see Ref.~\cite{EPAPS}). Despite these comparable FM moments, these SLs have remarkably different magnetic depth profiles. This is evident from the marked difference of the PNR curves in the vicinity of the 2$^{\text{nd}}$ and 4$^{\text{th}}$ order SLBPs. The even order SLBPs should have a very small intensity since they are symmetry forbidden due to the similar thicknesses of about 10\,nm of the YBCO and LMO or LCMO layers. The intensities of the 2$^{\text{nd}}$ and 4$^{\text{th}}$ SLBP are indeed strongly suppressed for the curves at 300\,K where only the nuclear potentials contribute.

\begin{figure}
\includegraphics[width=0.6\textwidth]{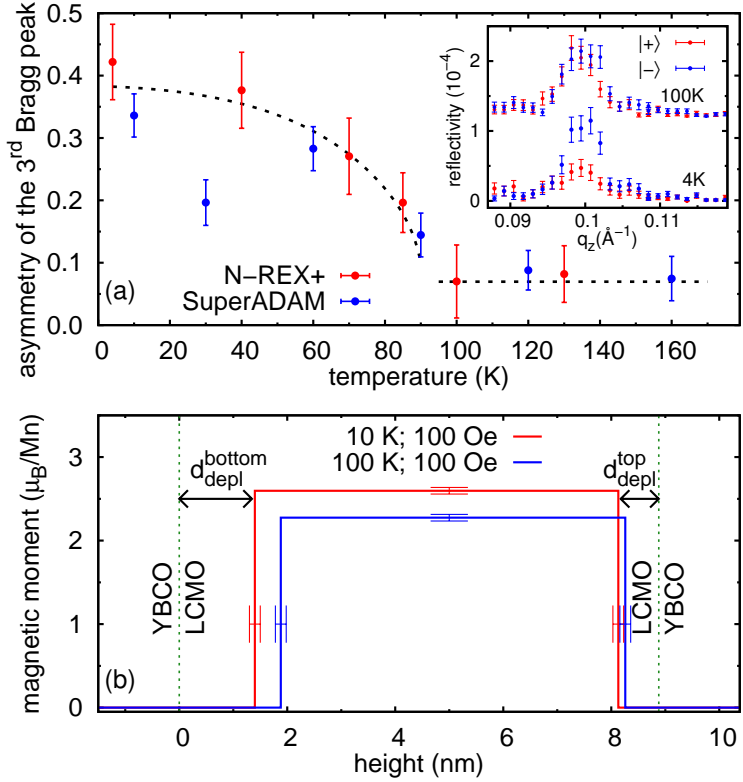}
\caption{\textbf{a)} $T$-dependence of the asymmetry between the peak intensities of the spin-down and spin-up polarized curves at the 3$^\text{rd}$ order SLBP showing an anomalous order-parameter-like increase below $T_{\text{c}}$. Inset: Magnification around this peak above and well below $T_{\text{c}}$. \textbf{b)} Corresponding changes of the magnetic depth profile.\label{Fig2}}
\end{figure}

For the SL with the FM-I LMO layers in Fig. \ref{Fig1}a the intensity of the 2$^{\text{nd}}$ and 4$^{\text{th}}$ order SLBPs remains weak even at  $T\ll T^{\text{Curie}}$. This confirms that the magnetic potential essentially maintains the symmetry of the nuclear one, i.e. that the magnetization is not strongly modified by a MPE. The magnetic depth profile in Fig.~\ref{Fig1}b obtained from fitting the data in Fig.~\ref{Fig1}a (solid lines) shows indeed that the FM order persist throughout the LMO layers and disappears close to the LMO/YBCO interfaces, i.e. within the combined error bar of the nuclear and magnetic potential widths of about $1.5\,$\AA. In clear contrast, for the FM-M LCMO a pronounced  2$^{\text{nd}}$ order SLBP develops below $T^{\text{Curie}}$. This highlights that the profile of the magnetic potential is substantially different from the nuclear one. This effect was previously noted~\cite{Stahn_2005,Hoffmann_2005,Chakhalian_2006} and interpreted in terms of two possible magnetic depth profiles. The so-called inverse proximity effect model assumes an induced Cu moment in the YBCO layers that is antiparallel to the Mn one. Alternatively, the dead layer or better depleted layer model (as we argue below) involves a suppression of the FM order of the Mn moment on the LCMO side of the interface~\cite{PRL_Luo_2004,PRB_Przyslupski_2004}. The previous PNR data contained only the 1$^\text{st}$ and 2$^\text{nd}$ order SLBPs which made it difficult to distinguish between these two possibilities. The accessible range of scattering vectors was limited by the quality of the SLs and, in particular, by the buckling of the SrTiO$_3$ substrates which develops below 100\,K in the context of a structural transition~\cite{Hoppler_2009,Hoppler_2008}. This problem has been avoided for our new SLs on LSAT substrates~\cite{Malik}. Our PNR curves extend with a sufficient signal to noise ratio to the $4^\text{th}$ order SLBPs and  thus provide the additional information that is required for an unambiguous identification of the magnetic depth profile. In particular, as shown in Fig.~\ref{Fig1}d, they confirm the depleted layer model where the FM order of the Mn moments is strongly reduced on the LCMO side of the interface. In these fits we neglected the induced FM Cu moments in the YBCO layers that are addressed below. They are much smaller than the Mn moments which thus govern the main features of the PNR curves. Interestingly, the depleted FM layers are significantly thinner at the top interface than at the bottom one as seen from the film surface. Whether this effect is caused by a difference of the atomic layer sequence or the interface roughness remains to be further investigated. Notably, the magnetic potentials at 10 and 100\,K as shown in  Fig.~\ref{Fig2}b exhibit  some characteristic differences, i.e. the value of $d_{\text{depl}}^{\text{bottom}}$  is reduced by about 5\AA\ where as $d_{\text{depl}}^{\text{top}}$ is hardly changed. A more detailed T-dependent study has been performed for the vicinity of the  3$^\text{rd}$ order SLBP where the changes between the curves at 10\,K and 100\,K are most pronounced.   Figure~\ref{Fig2}a, shows that the spin polarization dependent splitting of the intensity  of this SLBP exhibits an anomalous change in the vicinity of the SC transition of YBCO. It remains very small above 90\,K  and exhibits an order parameter like increase below  $T_{\text{c}}^{\text{ons}}\approx 88$\,K. A corresponding set of complete PNR curves would be required to establish whether this SC-induced anomaly involves a changes of the depleted layer thickness or rather a modification of the magnetic roughness, for example due to a changes in the FM domain structure. Irrespective of this open question, the mere observation of this SC induced anomaly implies that the coupling between the SC order in the YBCO layers and the magnetic one in LCMO is maintained despite of the depleted layer.  The later thus should not be mistaken for a magnetically dead layer but may host an antiferromagnetic or oscillatory component which cannot be detected with the specular PNR. The comparison between the LMO and LCMO SLs certainly points towards an intrinsic MPE that is intimately related to the electronic (orbital) properties of the FM manganite layers (more evidence is presented in Ref.~\cite{EPAPS}). 

\begin{figure}[b]
\includegraphics[width=0.5\textwidth]{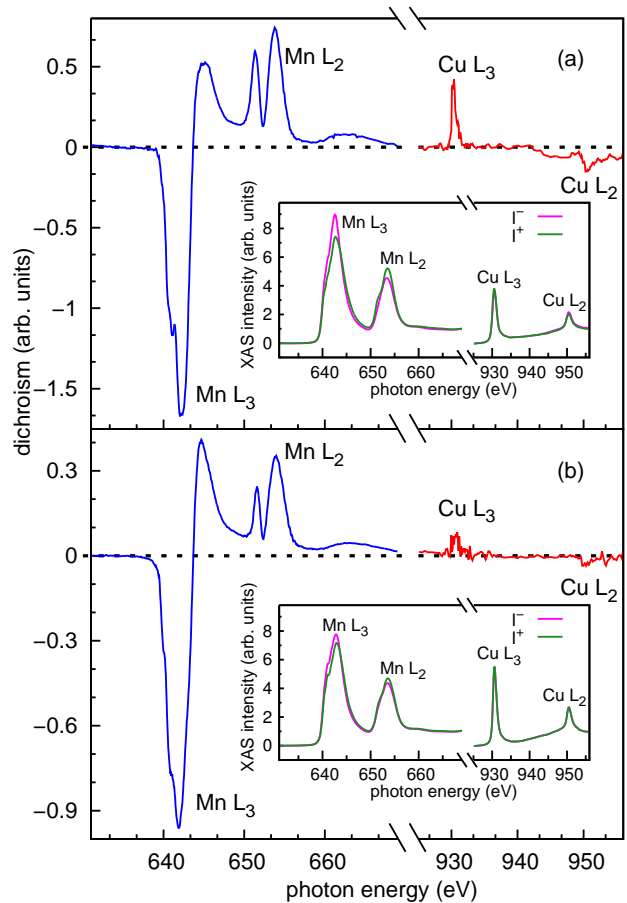}
\caption{XMCD spectra in TEY mode at the L-edges of Mn and Cu plotted in terms of the dichroism, $I^+-I^-$, for \textbf{a)} YBCO/LCMO and \textbf{b)} YBCO/LMO. Inset: Corresponding normalized x-ray absorption spectra.  \label{Fig3}}
\end{figure}

Further insight into the MPE has been obtained with XMCD measurements which provide element specific magnetic information. As was previously noted~\cite{Chakhalian_2006}, they reveal a FM moment of the Cu ions. Our new observation is that this Cu moment is significantly larger for FM-M LCMO than for the FM-I LMO. Figure 3 shows the XMCD difference spectra around the L$_3$ and L$_2$ edges of Mn and Cu as obtained in total electron yield (TEY) mode. For YBCO/LCMO in Fig.~\ref{Fig3}a they reveal a sizeable  average Cu moment per electron hole in the Cu 3d shell with an  spin component of 0.23$\,\pm\,0.1\,\mu_B/n_h$ (the orbital one is much smaller). For YBCO/LMO in Fig.~\ref{Fig3}b it amounts only to about 0.03$\,\pm\,0.01\,\mu_B/n_h$. The orientation of the Cu moment with respect to the Mn one is in both cases antiparallel as is evident from the opposite sign of the XMCD signals at the L$_3$ and L$_2$ edges of Cu and Mn. This Cu moment was previously explained as a consequence of antiferromagnetic superexchange interactions across the interface~\cite{Chakhalian_2006}. Our new PNR and XMCD data establish that these induced Cu moments exist despite the depleted FM layer on the LCMO side of the interface. Notably, the induced Cu moment is even significantly smaller for the YBCO/LMO SL where a large FM Mn moment persists to the interface. Once more this suggests that the depleted layers host a strong magnetic order that cannot be detected with PNR and XMCD (which are only sensitive to FM components). The induced Cu moment on the YBCO side and the strong modification of the FM order of the Mn moments on the LCMO side of the interface are likely caused by the same mechanism, a possible scenario in terms of a covalent bonding across the interface that involves the 3$d_{z^2-r^2}$ orbitals of Cu and Mn is outlined in Ref. ~\cite{Chakhalian_2007}. The strong suppression of this MPE for the FM-I LMO may well be related to the formation of orbital polaron arrays and other orbitally ordered states~\cite{Geck_2004} that start to compete with the interfacial covalent bonding as the manganite layers become less hole doped and insulating.
\begin{figure}[b]
\includegraphics[width=0.5\textwidth]{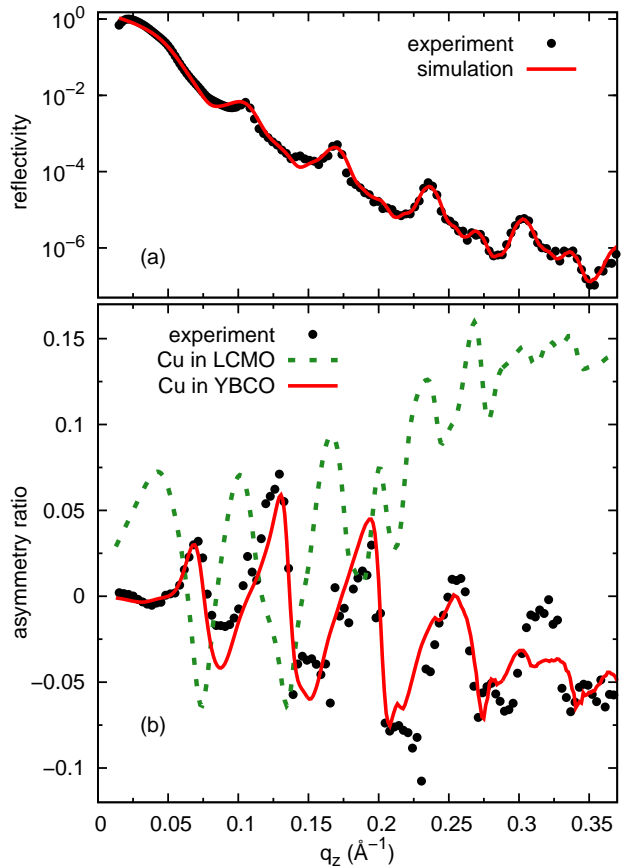}
\caption{\textbf{(a)} Resonant x-ray reflectometry curve at the Cu~L$_3$ edge of YBCO/LCMO obtained at 300\,K. \textbf{(b)} Difference curve between positive and negative x-ray helicity at 40\,K containing the information about the depth profile of the FM Cu moments. The simulations have been performed for the cases where the Cu moments reside in the YBCO layers (red line) and the LCMO layers (green dashed line), respectively.\label{Fig4}}
\end{figure}

Finally, we address the concern that the signal of the FM moments in the Cu-XMCD data may arise from a small number of Cu ions that are unintentionally incorporated within the FM manganite layers, for example due to chemical interdiffusion or another kind of cross contamination during PLD growth. The XMCD technique in TEY mode is indeed very surface sensitive and does not probe the spatial distribution of the magnetic moments. Therefore, we have also performed x-ray resonant magnetic reflectometry (XRMR) measurements near the Cu L$_3$ edge which allow one to probe how the Cu ions and their respective magnetic moments are distributed throughout the entire SL. The unpolarized reflectivity curve at 300\,K in Fig.~\ref{Fig4}a contains information about the depth profile of the concentration of the Cu ions. The red line shows the result of a simulation with a simple block-like depth profile of the complex chemical potentials (see Ref.~\cite{EPAPS}) which reproduces the period and the intensity of the main peaks and confirms the $\sim20$\,nm thickness of the YBCO/LCMO bilayers. The information about the corresponding magnetic depth profile of the Cu moments is contained in the asymmetry (difference) curve between positive and negative x-ray helicity at 40\,K as shown in Fig.~\ref{Fig4}b. This asymmetry curve contains a series of pronounced and equally spaced peaks which highlight that the FM Cu moments are periodically distributed throughout the SL. The red line shows the result of a simulation where the FM Cu moments are placed within the YBCO layers (for simplicity we also used block-like magnetic profiles). Based on the XMCD data we assume that the Cu moments are antiparallel to the Mn moments whose direction is determined by the external magnetic field and we use a Cu moment of about 0.25$\,\pm 0.1\,\mu_B$. While the fit could be further improved, for example by grading the profile of the Cu moments in the YBCO layers, it already reproduces quite well the main features of the data like the peak positions and their intensity variation. Most importantly, the simulation for the opposite case where the Cu moments are assumed to reside within the LCMO layers (dashed green line) is in disagreement with the data. The Cu moments are now placed on the opposite side of the interface which gives rise to a $\pi$ phase shift of the reflected x-ray waves and thus to an exchange of the maxima and minima of the asymmetry curve. A reasonable agreement with the data could only be obtained by changing the mutual orientation of the Cu and Mn moments from antiparallel to parallel. This possibility is however excluded by the XMCD data in  Fig.~\ref{Fig3} which highlight that the Cu and Mn moments are antiparallel. Our combined XMCD and XRMR data thus provide compelling evidence that the FM Cu ions originate from the YBCO layers.

In summary, we performed polarized neutron reflectometry, x-ray circular dichroism and  x-ray resonant magnetic reflectometry measurements on YBCO/LCMO and YBCO/LMO SLs whose FM manganite layers are metallic and insulating, respectively. We found that the MPE is governed by the electronic (orbital) state of the FM manganite layers. Furthermore, we detailed the MPE induced changes of the magnetic depth profile and provided direct evidence that the induced Cu moments reside within the YBCO layers. Our results thus provide evidence for an intrinsic nature of the MPE and an intimate magnetic coupling across the cuprate/manganite interfaces.
\begin{acknowledgments}
The UniFr group was supported by the SNF grants 200020-11978 and 200020-129484 and the NCCR MaNEP, the MPI-FKF group by the DFG Grant No. SFB/TRR 80. This work is partially based on experiments performed on Amor at the Swiss spallation neutron source SINQ, PSI, Villigen, CH, on NREX at FRM II, Munich, D, on SUPERADAM at ILL, Grenoble, F, on UE56/2-PGM1 at BESSY II, Berlin, D and on WERA at ANKA, Karlsruhe, D. 
\end{acknowledgments}

%% The bibilography


\bibliography{biblio_short}
\end{document}